\documentclass[review]{elsarticle}

\usepackage{lineno,hyperref}

\usepackage[margin=1in]{geometry}

\pdfoutput=1

\usepackage{graphicx}
\usepackage{bm}
\usepackage{dcolumn}
\usepackage{setspace}
\usepackage{amsmath}
\usepackage{fancyhdr}

\usepackage{caption}
\usepackage{subcaption}

\usepackage{color,soul}

\modulolinenumbers[5]

\journal{Computational Materials Science}

\makeatletter
\AtBeginDocument{\let\hl\@firstofone}
\makeatother

\begin{document}

\begin{frontmatter}

\title{Twin interaction with $\Sigma11$ tilt grain boundaries in BCC Fe : Formation of new grain boundaries}


\author[MDTD]{G. Sainath\corref{mycorrespondingauthor}}
\cortext[mycorrespondingauthor]{Corresponding author}
\ead{sg@igcar.gov.in}

\author[MDTD]{A. Nagesha}

\address[MDTD]{Materials Development and Technology Division, Metallurgy and Materials Group, Indira Gandhi Centre for Atomic Research, HBNI, Kalpakkam, Tamilnadu-603102, India}


\begin{abstract}

It is well known that the twinning is an important mode of plastic deformation in nanocrystalline materials. As a result, it is 
expected that the twin can interact with different grain boundaries (GBs) during the plastic deformation. Understanding these 
twin-GB interactions is crucial for our understanding of mechanical behavior of materials. In this work, \hl{the twin interaction 
with different $\Sigma11$ symmetric and asymmetric tilt GBs has been investigated in BCC Fe using molecular dynamics (MD) simulations.
The results indicate that twin nucleate from the crack or GB and, its interaction with $\Sigma11$ asymmetric tilt GBs leads to the 
formation of a new GB}. This \hl{new GB} consist of $<$100$>$ Cottrell type immobile dislocations. The detailed atomistic mechanisms 
responsible for this \hl{new GB} formation have been revealed using atomistic simulations. \hl{Interestingly, the new GB formation 
has not been observed in the case of twin interaction with $\Sigma11$ symmetric tilt GBs}. \\
\end{abstract}


\end{frontmatter}

{\centering

\section*{Highlights}

\begin{itemize}

\item In BCC Fe, interaction of twin with $\Sigma11$ ATGBs has been investigated using MD simulations.

\item Twin interaction can transform the $\Sigma11$ ATGB into a new GBs consisting of [100] type immobile dislocations.

\item The detailed atomistic mechanisms responsible for the new GB formations have been revealed.

\item The new GB formation occurred according to $\Sigma3 + \Sigma11 \longrightarrow \Sigma33$ and $\Sigma11 \longrightarrow \Sigma3 + \Sigma33$ reactions.

\end{itemize}
}

\pagebreak

{\centering
\section*{Graphical Abstract}
 \begin{figure}[h]
 \centering
 \includegraphics[width=14cm]{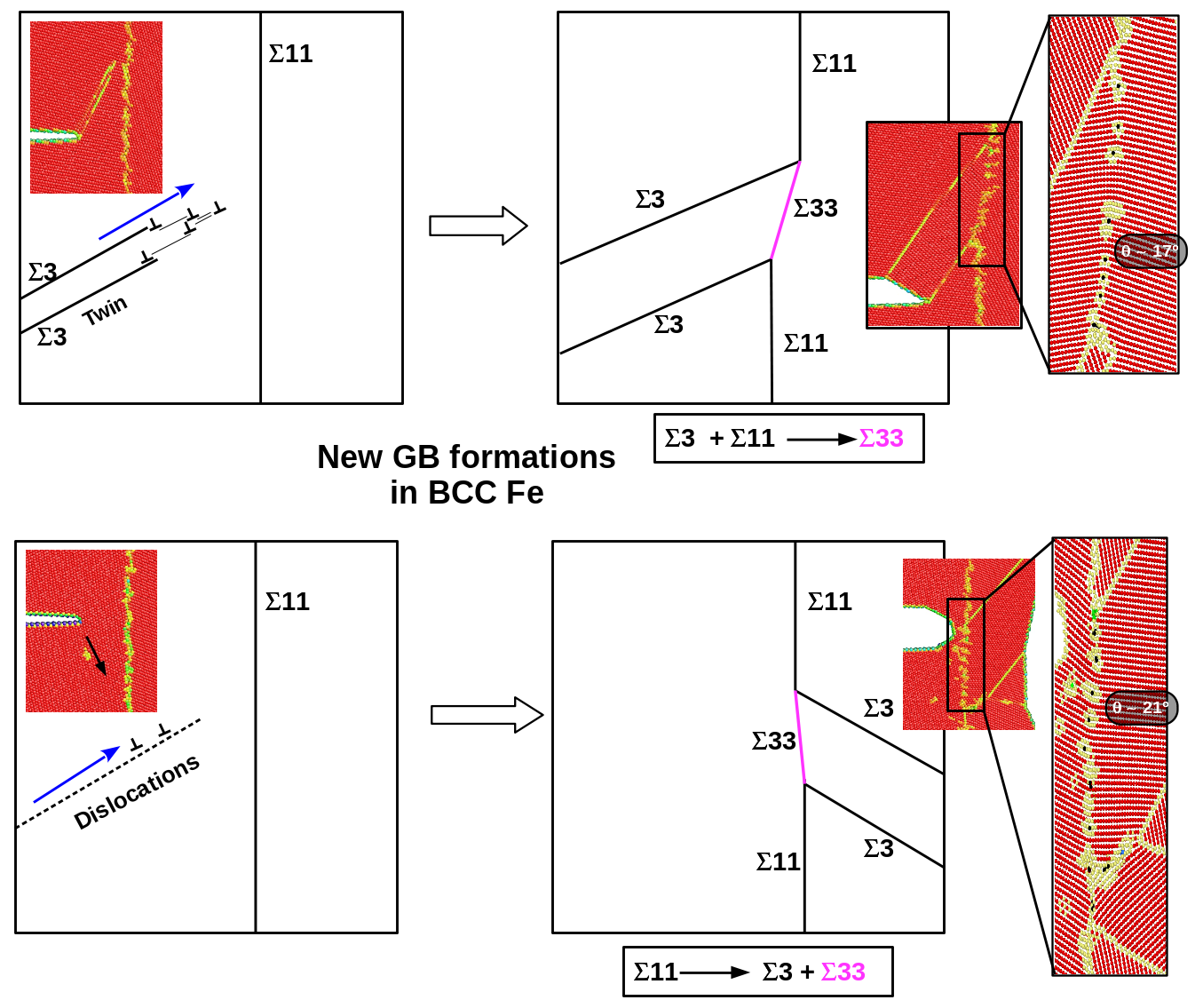}
 \label{abstract}
 \end{figure}
 }


\section{Introduction}
Many structural materials are generally polycrystalline in nature and thus contain grain boundaries (GBs) with different structures 
and wide spectrum of energies \cite{Randle-25,Watanabe}. The structure and energy of the GBs determine the physical, chemical and 
mechanical properties of polycrystalline materials. Especially, the GBs play an important role in plastic deformation, radiation 
damage, corrosion resistance, crack propagation resistance and failure \cite{Randle-25,Watanabe,Randle-139,Chinese,Sainath-INAE}. 
During the plastic deformation, the GBs interact with nearby grain boundaries, dislocations, twins, cracks, precipitates and many 
other defects. These defect interactions in turn lead to the grain/GB refinement, thus making the deformation more complex. As a 
result, investigating the defect-defect interactions through conventional experiments becomes difficult. In this context, the 
atomistic simulations can be utilized as an effective and efficient tool to examine the defect-defect interactions during the 
plastic deformation.

In the past many studies have been performed to understand the grain refinement through dislocation-GB interactions. For example, 
Koning et al. \cite{Koning} have shown that the dislocation transmission across the $\Sigma 11$ symmetric tilt GB results in 
local GB migration and disconnection formation in Ni. This process involves the motion of GB dislocations, which are created due 
to the partial absorption of incoming dislocations. In a similar fashion, a twin boundary (TB) can transform itself into a curved 
GB due to the continuous impingement and accumulation of high density of dislocations \cite{NTRao}. This process occurs due to the 
rearrangement of densely arranged dislocations, which minimizes the strain energy. It has been shown that in nanocrystalline Pt, the 
glide of partial dislocations on two intersecting \{111\} planes transform the original $<$110$>$ $\Sigma$3 symmetric tilt GB into a 
$<$110$>$ $\Sigma$9 symmetric tilt high angle GB \cite{Ultramicroscopy}. This TB transformation process has accommodated as large as 
47\% shear strain without initiating any cracks. In addition to dislocation-GB interactions, grain refinement via deformation 
twinning/de-twinning has also been reported. In duplex stainless steels, Cao et al. \cite{Cao-detwinning} have observed the formation 
of a new low angle GB or wall of dislocations due to the de-twinning of nanotwins. Similarly, the deformation twinning associated partial 
dislocation activity in Cu nanopillars has transformed the coherent TB into $\Sigma$9 boundary and also resulted in the formation of 
five-fold twin \cite{Sai-CMS}. This transformation occurs when the edge of a micro twin approaches towards an already existing twin 
leading to the formation of $\Sigma$9 boundary at their intersection. \hl{The twin interaction with various twist and tilt GBs in HCP 
Mg has revealed that all the GBs were observed to have a blocking effect on deformation twinning and no new GB formation has been 
observed \mbox{\cite{HCP-Mg-IJP}}}. 
Apart from transforming the GB, the defect-GB interactions can also lead to annihilation of the GB itself. For example, the de-twinning 
in Cu and Fe nanopillars has led to the annihilation of TBs 
itself, thus making the twinned grains larger in size \cite{PRL,Sai-crystals}. Thus, it is important to characterize the dislocation-GB, 
twin-GB and crack-GB interactions in order to understand the grain refinement and GB engineering processes in materials. 

However, most 
of these studies were focused on symmetric tilt GBs in FCC materials, even though the GBs in real polycrystalline materials are mostly 
asymmetric. \hl{Very few studies exist in BCC systems like Fe, which forms a basis for different structural materials of nuclear 
reactors. In BCC Fe, the twin interaction with $\Sigma$3 boundaries has been investigated using atomistic simulations \mbox{\cite{Sai-crystals}}. 
It has been shown that the twin penetrates across the $\Sigma$3 boundary in two different ways; (1) it can penetrate directly to the 
next grain without any deviation in twinning plane and (2) it can pass onto a plane symmetrical to the original twinning plane. Further, 
this penetration has led the complete annihilation of pre-existing TBs \mbox{\cite{Sai-crystals}}. However, in order to develop more 
reliable GB strengthening models, there is a need to study the 
twin interactions with many other grain boundaries like $\Sigma$5, $\Sigma$9, $\Sigma$11 etc. In view of this, the present study is 
focused on understanding the twin interaction with $\Sigma$11 symmetric and asymmetric tilt GBs in BCC Fe}.

\section{Simulation Methodology}

\subsection{Creation of grain boundaries}

\begin{figure}[h]
\centering
\includegraphics[width=16cm]{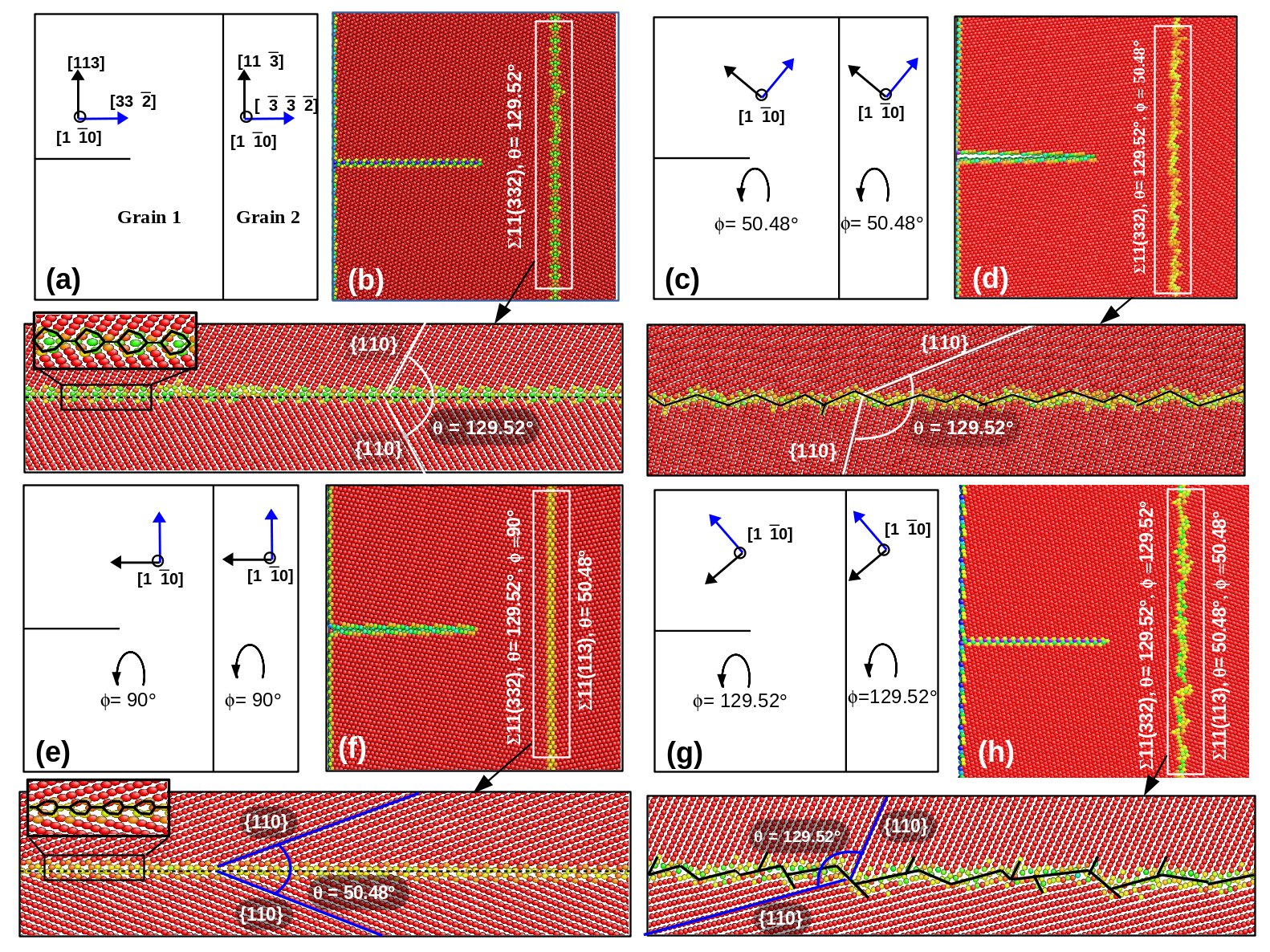}
\caption{\hl{A schematic of the process of creating $\Sigma$11 symmetric and asymmetric tilt GBs along with their structures in BCC Fe. (a,b) 
$\Sigma$11(332) STGB, (c,d) $\Sigma$11(332) ATGB with $\phi = 50.48^o$, (e,f) $\Sigma$11(113) STGB, and (g,h) $\Sigma$11(332) ATGB with 
$\phi = 129.52^o$ or $\Sigma$11(113) ATGB with $\phi = 50.48^o$. The atoms in (b,d,f and h) are colored according to their centro-symmetry 
parameter (CSP) and visualized using AtomEye.}}
\label{Initial}
\end{figure}

Molecular dynamics (MD) simulations have been carried out to study the interaction of a twin and crack with $\Sigma$11 \hl{symmetric 
and} asymmetric tilt GBs in BCC Fe. \hl{Two different $\Sigma$11 symmetric and asymmetric tilt GBs have been considered in this study. 
These GBs have been created as follows; first, the $\Sigma$11(332) symmetric tilt grain boundary (STGB) has been} created by joining 
two separate \hl{grains, 1 and 2}, with crystallographic orientations as shown in Fig. \ref{Initial}a. This joining results in 
the formation of $\Sigma$11(332)$[1\bar10]$ STGB with a misorientation angle ($\theta$) of $129.52^o$ at the interface (Fig. 
\ref{Initial}b). Following this, both \hl{grain 1 and 2}, were rotated with an equal angle of $\phi$, known as GB inclination angle, 
around the $[1\bar10]$ axis (Figs. \ref{Initial}c,e,g). The rotation \hl{of grain 1 and 2} with an inclination angle ($\phi$) = $50.48^o$ 
as shown in Fig. \ref{Initial}c results in the formation of $\Sigma$11(332) asymmetric tilt grain boundary (ATGB) with $\theta$ = $129.52^o$ 
and $\phi$ = $50.48^o$ at the interface (Fig. 
\ref{Initial}d). \hl{Similarly, the rotation with $\phi$ = $90^o$ as shown in \mbox{Fig. \ref{Initial}e} leads to the formation of 
$\Sigma$11(113) STGB with a misorientation angle ($\theta$) = $50.48^o$ at the interface \mbox{(Fig. \ref{Initial}f)}. This boundary 
can also be represented as $\Sigma$11(332) with $\theta$ = $129.52^o$ and $\phi$ = $90^o$. Finally, the rotation with $\phi$ = $129.52^o$ 
(\mbox{Fig. \ref{Initial}g}) results in the formation of $\Sigma$11(332) ATGB with $\theta$ = $129.52^o$ and $\phi$ = $129.52^o$ at 
the interface \mbox{(Fig. \ref{Initial}h)}. This boundary can also be represented as $\Sigma$11(113) ATGB with $\theta$ = $50.48^o$ and 
$\phi$ = $50.48^o$ \mbox{(Fig. \ref{Initial}h)}. This procedure of creating symmetric and asymmetric GBs is similar to that adopted 
in \mbox{Ref.\cite{JAP-GB-creation}}.}

\subsection{Structure of grain boundaries}

\hl{The detailed analysis of the structure of the GBs indicated} that the STGBs lie on single plane, (332) (Fig. \ref{Initial}b) and (113) 
(Fig. \ref{Initial}f) and these GB planes are inclined symmetrically with respect to \{110\} planes (Inset of Fig. \ref{Initial}b and f). 
Further, STGBs consist of periodically arranged structural units. On the contrary, ATGBs possesses an asymmetric inclination with respect to the \{110\} 
planes and also the GB dissociates into small facets consisting of two different planes (Fig. \ref{Initial}d and h). This dissociation is in 
agreement with the fact that the ATGB would break up into small facets \cite{Asym-facets}.

\subsection{MD simulation details}

In order to study the interaction of twin with  $\Sigma$11 GBs, a sharp crack which acts as a twin \hl{or dislocation} nucleation site has 
been introduced perpendicular to the GB plane (Fig. \ref{Initial}a-h). \hl{This sharp crack has been created by turning off the interactions 
between two slabs (groups) of atoms, i.e., by excluding the interactions between the fracture surfaces. This method effectively creates a crack 
between the two slabs without removing any atoms}. The GB-crack system has a dimensions of $17.37 \times 34.10 \times 
1.61 $ nm along $x$-$<$332$>$, $y$-$<$113$>$, and $z$-$<$110$>$ directions. The crack length was half the width of the specimen and the GB 
is placed at distance of 4.35 nm from the crack tip. Periodic boundary conditions were applied only in the crack front direction, i.e., 
$<$110$>$. On this model system, the mode-I loading has been simulated using MD simulations. All the simulations were carried out in 
LAMMPS package \cite{Plimpton-1995} employing the Mendelev embedded atom method potential for BCC Fe \cite{Mendelev-Fe}. Before 
loading, energy minimization was performed by conjugate gradient method to obtain a relaxed structure of the GB. Velocity verlet 
algorithm was used to integrate the equations of motion with a time step of 2 fs. Following energy minimization, the model system 
was equilibrated to a required temperature of 10 K in a canonical ensemble. Upon thermalization, the deformation was carried out in 
a displacement-controlled manner at constant strain rate of $1 \times 10^8$ s$^{-1}$ by imposing displacements to atoms along the 
$y$-axis that varied linearly from zero at the bottom fixed layer to a maximum value at the top fixed layer. \hl{All the simulations 
were carried out till the strain reaches the value of 0.25}. The stress was obtained 
using the Virial expression, which is equivalent to a Cauchy's stress in an average sense. AtomEye package \cite{AtomEye}and OVITO 
\cite{Ovito} have been used for the visualization and dislocation analysis.

\section{Results}

\subsection{Twin interaction with $\Sigma$11(332) $\phi = 50.48^o$ ATGB}

\begin{figure}[h]
\centering
\includegraphics[width=16cm]{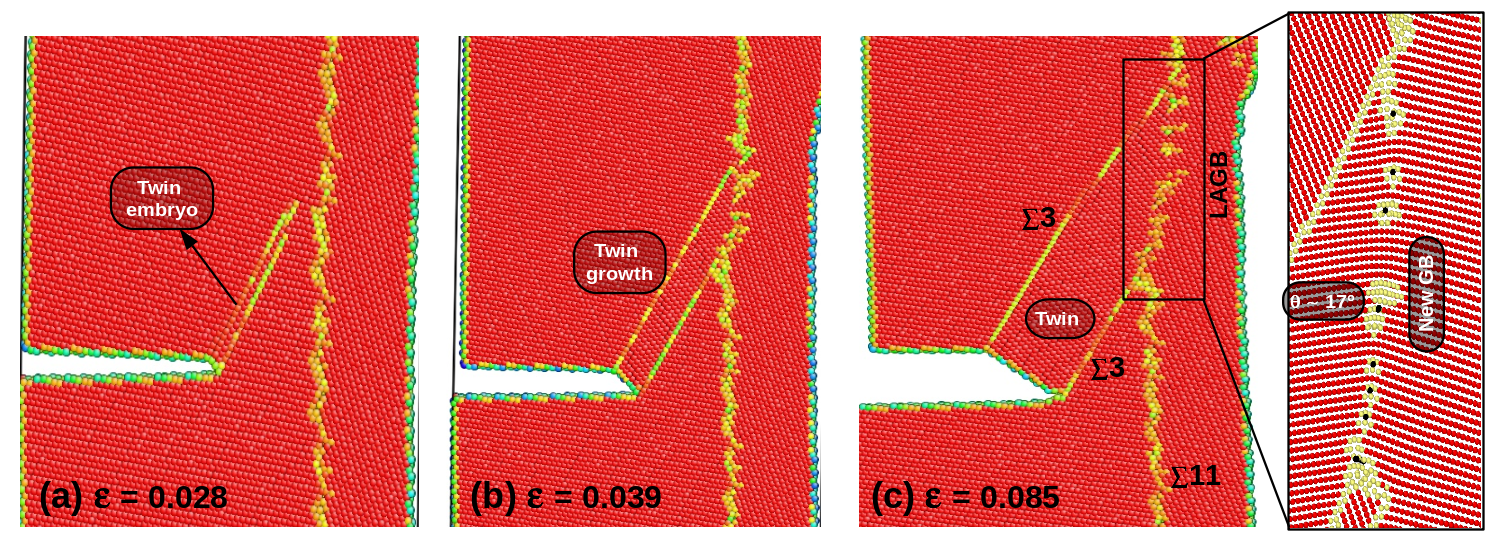}
\caption {Formation of \hl{a new GB} due to the interaction of a twin with \hl{$\Sigma$11(332) $\phi = 50.48^o$ ATGB} in BCC Fe. (a) The 
twin nucleation 
from a crack tip, (b) twin growth, (c) \hl{a new GB} formation at the interaction place of twin-ATGB. The atoms are colored according to their 
centro-symmetry parameter and visualized using AtomEye. The inset figure in (c) is visualized using OVITO. The black dots in inset of figure 
(c) are $<$100$>$ type dislocation lines.}
\label{yielding}
\end{figure}

Figure \ref{yielding} shows the formation of \hl{a new} GB due to the interaction of a twin \hl{(region enclosed by $\Sigma3$(112) 
TBs)} with $\Sigma$11\hl{(332) $\phi = 50.48^o$ 
ATGB} in BCC Fe. It can be seen that initially a twin embryo nucleates from the crack tip and approaches towards the ATGB (Fig. 
\ref{yielding}a). Once the twin front reaches the ATGB, its penetration and transmission to the neighboring grain is restricted (Fig. 
\ref{yielding}b). Following this, the width of the twin increases with increasing strain due to the continuous glide of twinning partials 
along the TBs (Fig. \ref{yielding}b-c). In this process of twin growth, the twinning partials interact with ATGB and transform the ATGB 
into \hl{a new GB} as shown in Fig. \ref{yielding}c). This \hl{new GB} consist of $<$100$>$ type immobile dislocations separated by a 
distance (d) ranging from 0.68 to 1.25 nm. The misorientation angle ($\theta$) across this boundary with respect to $<$110$>$ misorientation 
axis is found to be close to $17^o$, which makes it into the class of \hl{lower side of the medium angle GBs}. It has also been found that 
the mean separation distance between the GB dislocations obeys the well-known Frank-Bilby equation \cite{Frank-Bilby}, i.e. $d = b/(2sin(\theta/2))$, 
where b is the Burgers vector of GB dislocations.

\begin{figure}[h]
\centering
\includegraphics[width=16cm]{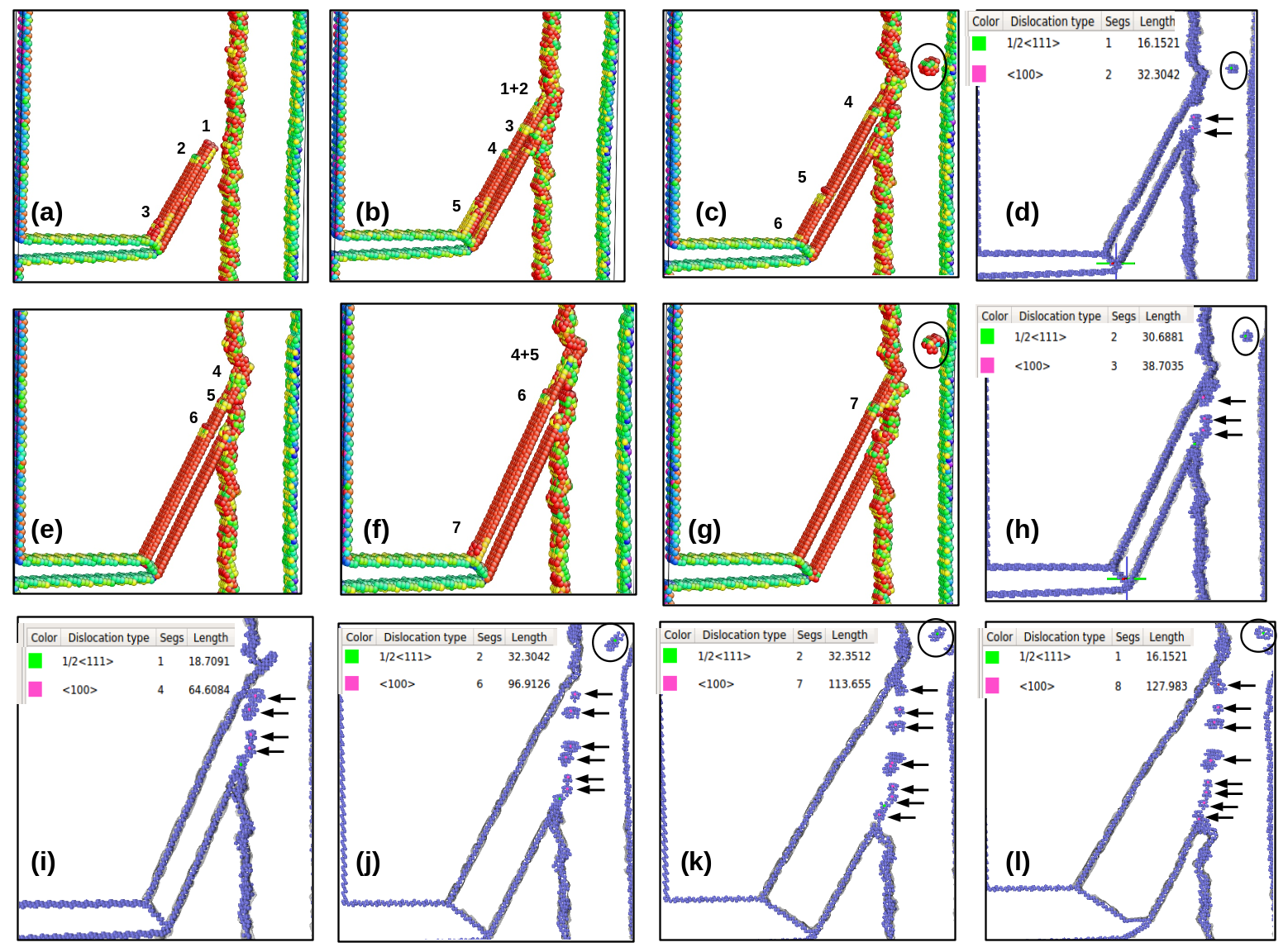}
\caption {The detailed atomistic and dislocation mechanism responsible for the transformation of $\Sigma$11(332) $\phi = 50.48^o$ ATGB into 
a \hl{new GB}. The atoms in figure (a-c) and (e-g) are colored according to their centro-symmetry parameter and visualized using 
AtomEye. The atoms in figure (d) and (h-l) are colored according to their CNA parameter and visualized using OVITO. \hl{The inset tables show 
dislocation analysis results using OVITO}. The magenta lines (dots in 2D) are $<$100$>$ type dislocations, while the green lines are 1/2$<$111$>$ 
type dislocations. \hl{The numbers 1,2,...7, represent the partial dislocations gliding on twin boundaries. For clarity, the $<$100$>$ 
type dislocations (pink dots) were also shown with black arrow marks}.}
\label{Mechanism}
\end{figure}

Understanding the detailed mechanism responsible for such formation of \hl{a new GB} is 
important in order to unravel the mechanism of grain refinement or reconfiguration of the GB network during the plastic deformation. 
Figure \ref{Mechanism} shows the detailed atomistic and dislocation mechanisms responsible for the transformation of $\Sigma$11 
ATGB into \hl{a new GB}. It can be seen that the twin embryo nucleated from the crack tip consists of partial dislocations identified 
as 1,2 and 3 in Fig. \ref{Mechanism}a. With increasing strain, the twin front reaches the ATGB along with an increase in 
twin width and more partial dislocations, 4 and 5 (Fig. \ref{Mechanism}b). It can also be seen that the ATGB restrict the 
transmission of partial dislocations as well as twin into the next grain. Due to this restriction, the partial dislocations 
1,2 and 3 pile-up and combine to form a 1/2[111] full dislocation at the ATGB. This dislocation reaction can be written as 
\begin{equation} 
\label{eq:1}
1/6 [111] + 1/6[111] + 1/6[111] \longrightarrow 1/2[111].
\end{equation} With further deformation, this full dislocation is emitted into the next grain by leaving a Cottrell dislocation 
\cite{Cottrell} having a Burgers vector (b) of [010] on the GB (Fig. \ref{Mechanism}c-d). This reaction can be written as 
\begin{equation}
\label{eq:2}
1/2 [111] \longrightarrow [010] + 1/2[1\bar11].
\end{equation} Energetically, the above reaction is not feasible, i.e. violates the Frank criterion. However, the presence of 
high stresses within the GB make it feasible \cite{GB-stress}. The same process described in Equation \ref{eq:1} and \ref{eq:2} occurs for the 
next set of partial dislocations 4,5 and 6 as shown in Fig. \ref{Mechanism}e-f. As a result, the emission of one more full 
dislocation can be seen in Fig. \ref{Mechanism}g and h. This emission of full dislocation leaves another Cottrell dislocation 
adjacent to the first one (Fig. \ref{Mechanism}h). With increasing strain, more and more Cottrell dislocations were added 
adjacent to each other (Fig. \ref{Mechanism}i-l). As a result, an increase in number of Cottrell dislocations from 3 in Fig. 
\ref{Mechanism}d (Magenta color lines) to 8 in Fig. \ref{Mechanism}l can be seen. These series of [100] Cottrell dislocations 
constitutes a \hl{new GB} as shown in Fig. \ref{yielding}c. Thus, the interaction of a growing twin with an ATGB results in the formation 
of a \hl{new GB}.

\subsection{Twin interaction with $\Sigma$11(113) $\phi = 50.48^o$ ATGB}

\begin{figure}[h]
\centering
\includegraphics[width=16cm]{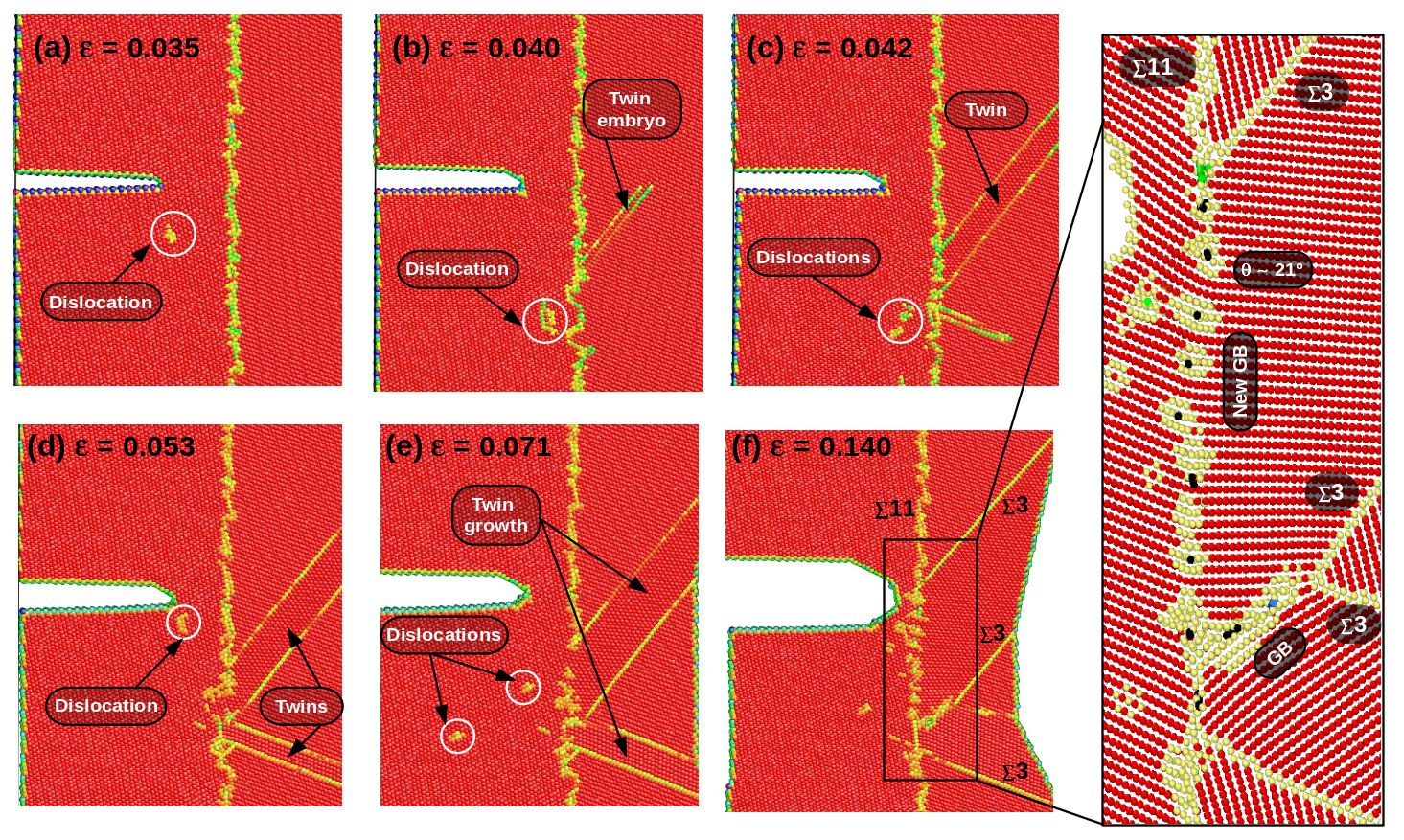}
\caption {\hl{Formation of a new GB due to the interaction of a dislocation/twin with $\Sigma$11(113) $\phi = 50.48^o$ ATGB in BCC Fe. (a) 
Dislocation nucleation from a crack tip, (b) nucleation of a twin from the GB due to dislocation-GB interaction, (c-d) twin growth along 
with new twin nucleation, (e) dislocation nucleation from the GB-twin junction back into the first/left grain, and (f) twin growth and new GB 
formation. The atoms are colored according to their centro-symmetry parameter and visualized using AtomEye. The inset figure in (f) is visualized 
using OVITO. The black dots in inset of figure (f) are $<$100$>$ type dislocation lines.}}
\label{Asymmetric-2}
\end{figure}

\hl{Figure \mbox{\ref{Asymmetric-2}} shows the formation of a new GB due to the interaction of a dislocation and twin with $\Sigma$11(113) 
$\phi = 50.48^o$ ATGB. It can be seen that, at first, a dislocation nucleates from the crack tip and glide towards the ATGB 
(\mbox{Fig. \ref{Asymmetric-2}a}). Once the dislocation reaches close to the boundary, it triggers the nucleation of twin from the ATGB
into the neighboring grain (\mbox{Fig. \ref{Asymmetric-2}b}). With increasing strain, more dislocations nucleate from the crack tip along 
with an increase in twin width (\mbox{Fig. \ref{Asymmetric-2}c}). Following this, one more twin nucleate from the ATGB and 
grow on a twin plane different from the original one \mbox{(Fig. \ref{Asymmetric-2}c-d)}. At this stage, the dislocations 
also start nucleating from the ATGB-twin junction into the first/left grain as shown in \mbox{Fig. \ref{Asymmetric-2}e}. The continuous 
growth of both the twins has led to the formation of a new GBs as shown in \mbox{Fig. \ref{Asymmetric-2}f} and inset. The careful examination of 
the newly formed GB has indicated that it consist of a series of $<$100$>$ type immobile dislocations separated by a distance (d) ranging 
from 0.6 to 1.0 nm and the misorientation angle ($\theta$) $\sim$ $21^o$ (lower side of the medium angle GBs) with respect to the $<$110$>$ 
misorientation axis. Also, the mean distance between the GB dislocations obeys Frank-Bilby equation \mbox{\cite{Frank-Bilby}}.}

\begin{figure}[h]
\centering
\includegraphics[width=16cm]{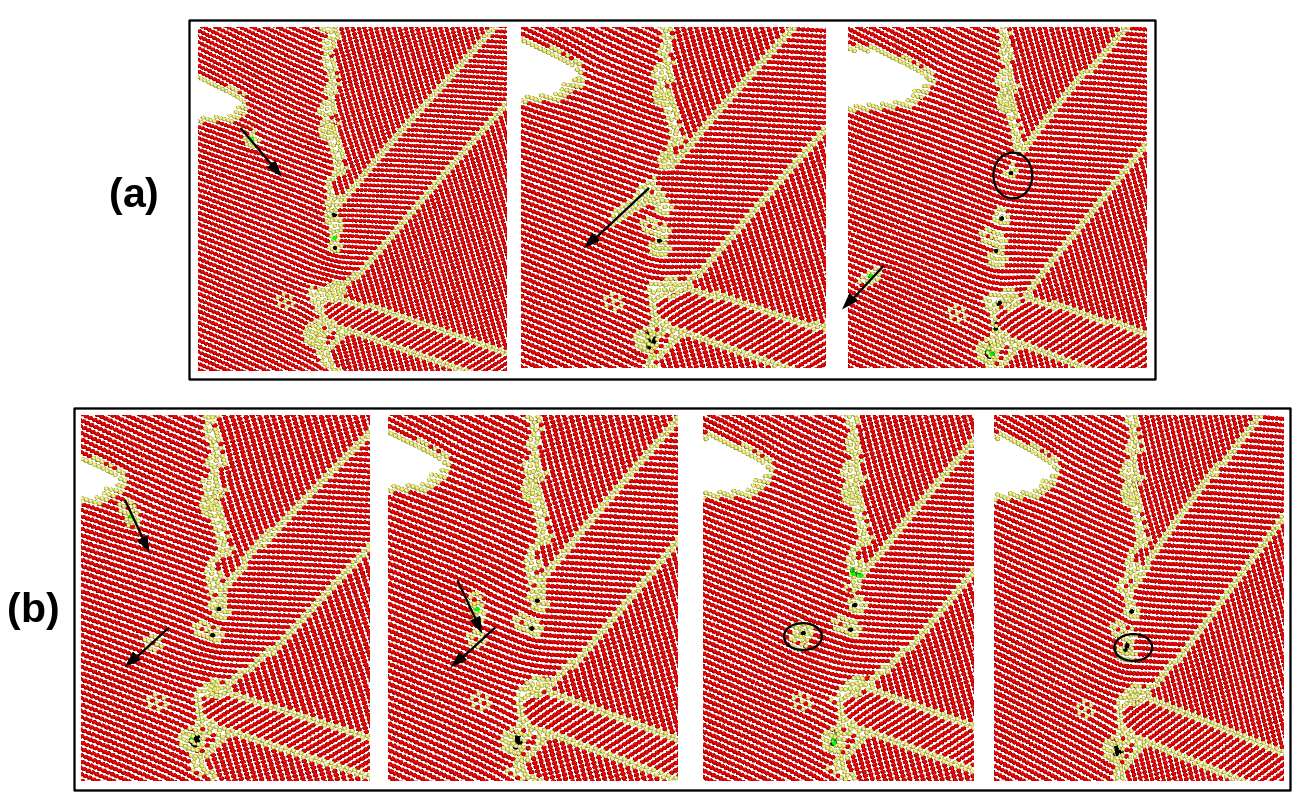}
\caption {\hl{The detailed atomistic and dislocation mechanism responsible for the transformation of $\Sigma$11(113) $\phi = 50.48^o$ ATGB into a 
new GB. Series of snapshots in (a) details a mechanism, where a dislocation nucleated from the crack tip is reflected by ATGB back into to the same 
grain by leaving a [100] dislocation on the new GB. Series of snapshots in (b) manifest a mechanism, where a dislocation nucleated from the crack tip 
interacts with the reflected dislocation and form a [100] type Cottrell dislocation slightly away from the boundary, which eventually attracted towards 
the new GB.}}
\label{Asymmetric-2-mech}
\end{figure}

\hl{\mbox{Figure \ref{Asymmetric-2-mech}} shows the detailed mechanisms responsible for the formation of a new GB in \mbox{Fig. \ref{Asymmetric-2}}. 
It has been observed that, following twin formation, the subsequent nucleation of dislocations from the crack tip is mainly responsible for 
both adding dislocations to new GB and also aiding the twin growth. The series of snapshots in \mbox{Figure \ref{Asymmetric-2-mech}a} show a 
mechanism where a dislocation nucleated from the crack tip is reflected by the $\Sigma$11(113) GB back into to the same grain by leaving a Cottrell 
dislocation \mbox{\cite{Cottrell}} with a Burgers vector of [100] on the GB. This reaction can be written as }\begin{equation}
\label{eq:3}
1/2 [\bar111] \longrightarrow [100] + 1/2[\bar1\bar1\bar1].
\end{equation} \hl{Here, $1/2 [\bar111]$ is the dislocation nucleated from the crack tip, $[100]$ is the Cottrell type GB dislocation and 
$1/2[\bar1\bar1\bar1]$ is the dislocation reflected back by the ATGB into the original grain \mbox{(Fig. \ref{Asymmetric-2-mech}a)}. Different 
from this, the snapshots in \mbox{Fig. \ref{Asymmetric-2-mech}b} shows another mechanism, where a dislocation nucleated from the crack tip 
($1/2[\bar111]$) interacts with the reflected dislocation from the earlier reaction \mbox{(Eq. \ref{eq:3})}, and form a $[100]$ type Cottrell 
dislocation slightly away from the GB according to the following reaction}; 
\begin{equation}
\label{eq:4}
1/2 [\bar111] + 1/2[\bar1\bar1\bar1] \longrightarrow  [100].
\end{equation} \hl{Eventually, this $[100]$ dislocation is attracted and readjusted into the GB \mbox{(Fig. \ref{Asymmetric-2-mech}b)}. These 
mechanisms shown in \mbox{Fig. \ref{Asymmetric-2-mech}a-b} occur multiple times leading to the addition of many $[100]$ type dislocations to the new GB.}

\subsection{Crack interaction with $\Sigma$11(332) and $\Sigma$11(113) STGBs}

\begin{figure}[h]
\centering
\includegraphics[width=16cm]{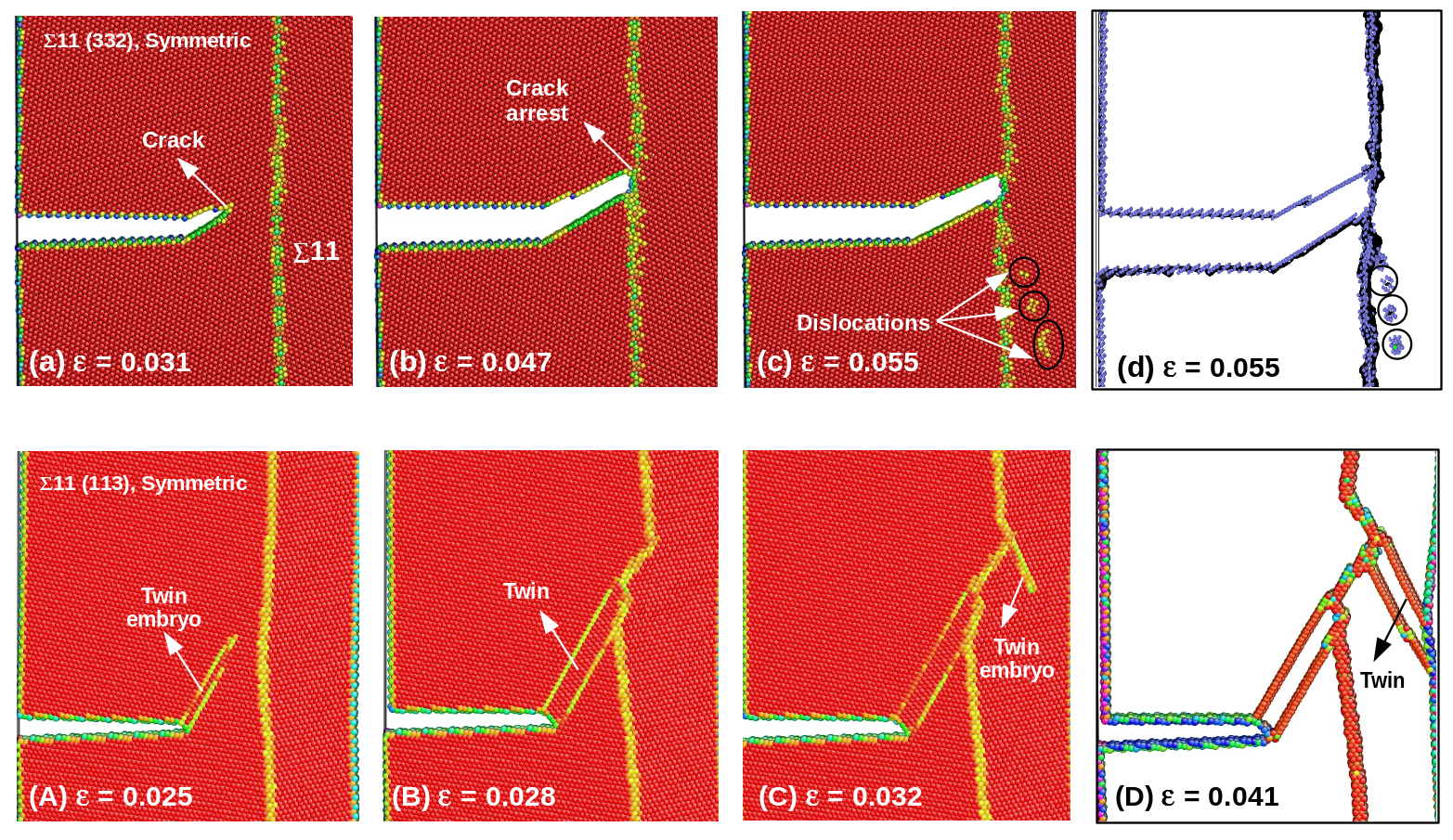}
\caption {\hl{The crack growth behavior in presence of (a-d) $\Sigma$11(332) STGB, and (A-D) $\Sigma$11(113) STGB in BCC Fe. In (a-d) once the 
brittle crack reaches the GB, the crack gets blunted significantly due to the emission of multiple dislocations from the GB and no transformation 
of GB is seen. In (A-D) twin nucleates from the crack tip and once it reach the GB, its transmission to the next grain is restricted leading 
to the distortion of STGB without any transformation. The atoms in (a-c) and (A-D) are colored according to their centro-symmetry parameter 
and visualized using AtomEye. The atoms in figure (d) are colored according to their CNA parameter and visualized using OVITO.}}
\label{Symmetric}
\end{figure}

Similar to ATGBs, the crack growth behavior in the presence of \hl{$\Sigma$11(332) and $\Sigma$11(113) STGBs} has also been investigated 
and presented in Fig. \ref{Symmetric}. It can be seen that \hl{the crack in case of $\Sigma$11(332) STGB} grows in a brittle manner on 
\{110\} plane without any micro-twin at the crack tip (Fig. \ref{Symmetric}a). Once the crack tip reaches the GB, the crack gets blunted 
due to the emission of dislocations (Fig. \ref{Symmetric}b-d). With increasing strain, more and more dislocations emit from the crack-GB 
intersection \hl{leading to significant crack blunting. On the other hand, the crack in presence of $\Sigma$11(113) STGB emits a twin 
embryo (Fig. \mbox{\ref{Symmetric}A)}, which becomes a full twin once it reach the GB (Fig. \mbox{\ref{Symmetric}B)}. Subsequently, the 
twin interacts with $\Sigma$11(113) STGB and its transmission into the next grain is restricted. As a result, the STGB near the interaction 
site is distorted from its original plane. With increasing strain, a new twin nucleate from the distorted GB (Fig. \mbox{\ref{Symmetric}C-D)}. 
Further deformation has lead to the growth of this newly formed twin without forming any new GB. These results indicate that in case of 
$\Sigma$11 STGBs neither the transformation nor the formation of any new GB occurs. Similar results were observed in the case of $\Sigma$3 
STGB in BCC Fe \mbox{\cite{Sainath-INAE}}, where the brittle crack transmit to next grain without changing the structure of the GB. These 
results on $\Sigma$11 and $\Sigma$3 boundaries} suggest that the STGBs are stable against the plastic deformation.

\section{Discussion}
\hl{The present results suggest that the twin interaction with $\Sigma$11 ATGBs lead to the formation of a new GBs. These new GBs have 
misorientation angle of $17^o$ and $21^o$, which falls in the lower side of the medium angle GB spectrum. Since these boundaries also 
consist of a well separated dislocations, they can be considered as low angle GBs. Accordingly, the new GBs observed in the present 
investigation are classified as low angle GBs. The generally accepted misorientation angle for transition from low to high angle GBs 
is in the range of $10-20^o$, depending on the material, misorientation axis, GB normal etc.} \cite{GBs-transition}. The low angle GBs 
form when multiple dislocations rearrange themselves into configurations of lower energy during the plastic deformation \cite{Dieter}. 
However, the low angle GB formation due to the interaction of a twin with ATGB as observed in the present investigation is interesting. 
The low angle GBs consisting of Cottrell type immobile dislocations have important consequences in mechanical properties, grain refinement 
processes, GB engineering and deformation behavior of materials. It is known that the low angle boundaries have lower GB energies than 
regular high angle boundaries and accordingly, they contribute to the thermal stability of microstructures and hence improve the mechanical 
properties of polycrystalline materials \cite{Ovidko}. The low angle GBs also act as strong and tunable barriers (tunable barrier allows the 
transmission of only selective dislocations based on the orientation relationship) for dislocation motion, thus contributing to hardening 
as well as softening of materials\cite{Tunable}. 

\begin{figure}[h]
\centering
\includegraphics[width=12cm]{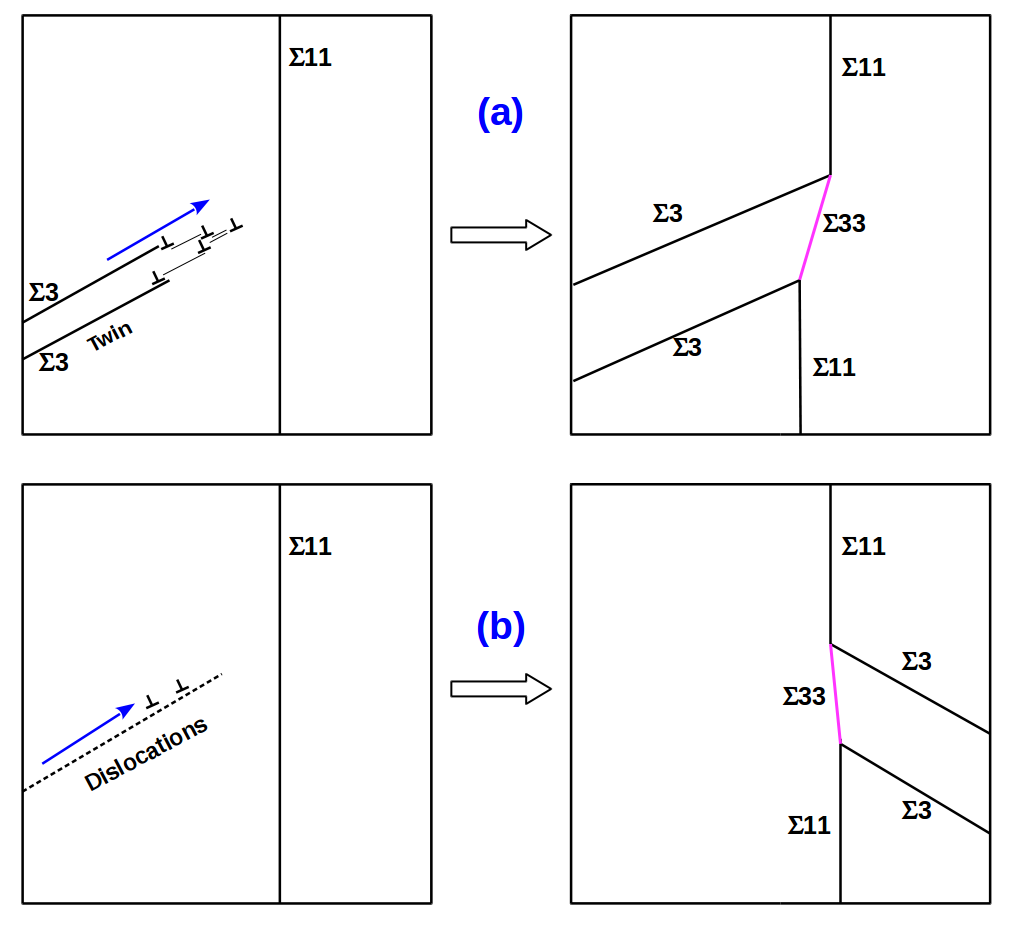}
\caption {A schematic of twin-ATGB interactions observed in the present study. (a) GB interactions observed in the case of $\Sigma11$(332) 
$\phi = 50.48^o$ ATGB ($\Sigma3 + \Sigma11 \longrightarrow \Sigma33$), and (b) GB interactions observed in the case of $\Sigma11$(113) 
$\phi = 50.48^o$ ATGB ($\Sigma11 \longrightarrow \Sigma3 + \Sigma33$).}
\label{Schematic}
\end{figure}

In GB engineering, it is established that the twinning alters and modifies mainly the $\Sigma$3 and its higher order boundaries i.e., 
$\Sigma$9 and $\Sigma$27, through the well known reaction 
\begin{equation}
\label{eq:5}
\Sigma3 + \Sigma3 \longrightarrow \Sigma3^n,
\end{equation} where n is an integer\cite{Randle-139}. However, the present results indicate that the twin can also alter the other 
high angle GBs like $\Sigma$11 and contribute to the formation of new GBs. \hl{As schematically shown in \mbox{Fig. \ref{Schematic}}, 
a possible reactions for this new GB formation can be written as} \cite{Randle-139,Humphreys}

\begin{equation}
\label{eq:6}
\Sigma3 + \Sigma11 \longrightarrow \Sigma33 \quad \text{(Figure } \ref{Schematic}\text{a)}
\end{equation} and 

\begin{equation}
\label{eq:7}
\Sigma11 \longrightarrow \Sigma3 + \Sigma33 \quad \text{(Figure } \ref{Schematic}\text{b)}.
\end{equation} \hl{Here, the $\Sigma$33 GB generally has a misorientation angle ($\theta$) = $20.1^o$ with respect to $<$110$>$ misorientation 
axis \mbox{\cite{Sigma33}}. Interestingly, the misorientation angles of new GBs observed in the present investigation are $\theta$ $\sim$ $17^o$ 
and $21^o$ with respect to the $<$110$>$ axis, which are close to misorientation angle of $\Sigma$33 GB. This suggest that the new GB observed 
in the present study is approximately $\Sigma33$ and its formation occurs due to the interaction of twin with $\Sigma11$ ATGB according to the 
\mbox{Eqs. \ref{eq:6} and \ref{eq:7}}. These kind of GB transformations strongly influence the structure and associated properties of many 
polycrystalline materials \mbox{\cite{Ovidko}}. However, the transformation of GBs has not been observed in the case of symmetric $\Sigma11$ 
GBs, indicating that the symmetric boundaries are stable against the GB transformations during the plastic deformation.}

\section{Conclusions}

The interaction of a twin with $\Sigma11$ \hl{symmetric and} asymmetric tilt GBs has been investigated in BCC Fe using atomistic 
simulations. The results indicated that the growing twin can transform the $\Sigma11$ ATGBs into \hl{a new GBs} consisting of [100] 
type immobile dislocations. This transformation \hl{has occurred} through a systematic activity of partial \hl{(1/6$<$111$>$)} and 
full \hl{(1/2$<$111$>$)} dislocations. The \hl{$\Sigma11$(332) $\phi=50.48^o$} ATGB restrict the penetration and transmission of a 
twin and twinning partials into the neighboring grain. Due to this restriction, the twinning partials \hl{(1/6$<$111$>$)} gliding 
on the TBs pile-up and combine to form a 1/2[111] full dislocation at the \hl{$\Sigma11$(332)} ATGB. With further deformation, this 
$1/2[111]$ full dislocation is emitted into the next grain by leaving a [100] Cottrell type dislocations on the GB. The same process 
occurs repeatedly leaving a series of [100] type immobile dislocations constituting a \hl{new GB. In case of $\Sigma11$(113) 
$\phi=50.48^o$ ATGB, once the dislocation nucleated from the crack tip reaches close to the boundary, it triggers the nucleation of 
twin from the boundary into the neighboring grain. Following twin formation, the subsequent nucleation of dislocations from the crack 
tip is mainly responsible for both adding dislocations to new GB and also aiding the twin growth. However, these kind of transformations 
or new GB formations have not been observed in the case of twin interaction with $\Sigma11$ symmetric boundaries}.

\hl{The misorientation angles of newly formed GBs in the present investigation are close to misorientation angle of $\Sigma$33 GB. This 
suggested that the new GB formation in the present study occurred according to $\Sigma3 + \Sigma11 \longrightarrow \Sigma33$ and $\Sigma11 
\longrightarrow \Sigma3 + \Sigma33$ reactions}. Thus, the present study sheds a light on the transformation mechanism of ATGB into 
new GBs due to deformation twinning and it has important consequences in grain refinement processes, grain boundary engineering and 
deformation behavior of materials. 

\section*{Data availability}

The data that support the findings in this paper are available from the corresponding author on request.


\section*{References}

\begin{thebibliography}{999}


\bibitem{Randle-25} V. Randle, Grain boundary engineering: an overview after 25 years, Mater. Sci. Tech. 26 (2010) 
253-261.

\bibitem{Watanabe} T. Watanabe, Grain boundary engineering: historical perspective and future prospects, J. Mater. 
Sci. 46 (2011) 4095-4115.

\bibitem{Randle-139} V. Randle, Twinning-related grain boundary engineering, Acta Mater. 52 (2004) 4067-4081.

\bibitem{Chinese} X. Xiao, H. Chu, H. Duan, Effect of grain boundary on the mechanical behaviors of irradiated 
metals: a review, Sci. China-Phys. Mech. Astron. 59 (2016) 664601.

\bibitem{Sainath-INAE} G. Sainath, A. Nagesha, Atomistic simulations of twin boundary effect on the crack growth 
behaviour in BCC Fe, Trans. Ind. Nat. Acad. Eng. (2021) 1-7.


\bibitem{Koning} M. de Koning, R.J. Kurtz, V.V. Bulatov, C.H. Henager, R.G. Hoagland, W. Cai, M. Nomura, Modeling of 
dislocation-grain boundary interactions in FCC metals, J. Nucl. Mater. 323 (2003) 281-289.

\bibitem{NTRao} N.R. Tao, K. Lu, Nanoscale structural refinement via deformation twinning in face-centered cubic metals, 
Scr. Mater. 60 (2009) 1039-1043.

\bibitem{Ultramicroscopy} L. Wang, T. Jiao, W. Yu, S. Xuechao, X. Sisi, M. Shengcheng, Y. Guanghua, Z. Zec, Z. Jind, H. 
Xiaodong, In situ atomic scale mechanisms of strain-induced twin boundary shear to high angle grain boundary in 
nanocrystalline Pt, Ultramicroscopy 195 (2018) 69-73.

\bibitem{Cao-detwinning} Y. Cao, Y.B. Wang, X.H. An, X.Z. Liao, M. Kawasaki, S.P. Ringer, T.G. Langdon, Y.T. Zhu, Grain boundary 
formation by remnant dislocations from the de-twinning of thin nano-twins, Scr. Mater. 100 (2015) 98-101.

\bibitem{Sai-CMS} G. Sainath, Sunil Goyal, A. Nagesha, Atomistic mechanisms of twin-twin interactions in Cu nanopillars, 
Comput. Mater. Sci. 185 (2020) 109950.

\bibitem{HCP-Mg-IJP} J. Tang, H. Fan, D. Wei, W. Jiang, Q. Wang, X. Tian, X. Zhang, Interaction between a $\{10\bar12\}$ twin 
boundary and grain boundaries in magnesium, Inter. J. Plast. 126 (2020) 102613.

\bibitem{PRL} G. Cheng, S. Yin, T.-H. Chang, G. Richter, H. Gao, Y. Zhu, Anomalous tensile detwinning in twinned nanowires. Phys. 
Rev. Lett. 119 (2017) 256101.

\bibitem{Sai-crystals} G. Sainath, Sunil Goyal, A. Nagesha, Plasticity through de-twinning in twinned BCC nanowires, Crystals 10 
(2020) 366.


\bibitem{JAP-GB-creation} X.J. Long, L. Wang, B. Li, J. Zhu, S.N. Luo, Shock-induced migration of $\Sigma$3$<$110$>$ grain boundaries 
in Cu, J. Appl. Phys. 121 (2017) 045904.


\bibitem{Asym-facets} D.L. Medlin, K. Hattar, J.A. Zimmerman, F. Abdeljawad, S.M. Foiles, Defect character at grain boundary facet 
junctions: Analysis of an asymmetric $\Sigma$5 grain boundary in Fe, Acta Mater. 124 (2017) 383-396.


\bibitem{Plimpton-1995} S. Plimpton, Fast parallel algorithms for short-range molecular dynamics, J. Comp. Phy. 117 (1995) 1-19.


\bibitem{Mendelev-Fe} M.I. Mendelev, S. Han, D.J. Srolovitz, G.J. Ackland, D.Y. Sun, M. Asta, Development of new interatomic 
potentials appropriate for crystalline and liquid iron, Philos. Mag. 83 (2003) 3977-3994.

\bibitem{AtomEye} J. Li, AtomEye: an efficient atomistic configuration viewer, Modell. Simul. Mater. Sci. Eng. 11 (2003) 173-177.

\bibitem{Ovito} A. Stukowski, Visualization and analysis of atomistic simulation data with OVITO-the Open Visualization Tool, 
Modell. Simul. Mater. Sci. Eng. 18 (2010) 015012.


\bibitem{Frank-Bilby} A.P. Sutton, R.W. Balluffi, Interfaces in Crystalline Materials, Oxford University Press, 1995.

\bibitem{Cottrell} A.H. Cottrell, Theory of brittle fracture in steel and similar metals, Trans. Met. Soc. AIME 212 (1958) 196.

\bibitem{GB-stress} Y. Guo, D.M. Collins, E. Tarleton, F. Hofmann, J. Tischler, W. Liu, R. Xu, A.J. Wilkinson, T.B. Britton, Measurements 
of stress fields near a grain boundary: Exploring blocked arrays of dislocations in 3D, Acta Mater. 96 (2015) 229-236.

\bibitem{GBs-transition} M. Winning, A.D. Rollett, Transition between low and high angle grain boundaries, Acta Mater. 53 (2005) 2901-2907.

\bibitem{Dieter} G. E. Dieter, Mechanical metallurgy, McGraw-hill, New-York, 1976.


\bibitem{Ovidko} S.V. Bobylev, M. Yu. Gutkin, I.A. Ovid’ko, Transformations of grain boundaries in deformed nanocrystalline 
materials, Acta Mater. 52 (2004) 3793-3805.

\bibitem{Tunable} S. Chen, Q. Yu, The role of low angle grain boundary in deformation of titanium and its size effect, 
Scripta Mater. 163 (2019) 148-151.


\bibitem{Humphreys} J. Humphreys and M. Hatherly, Recrystallization and related annealing phenomena, Elsevier, (1996) Page 217.

\bibitem{Sigma33} D. Scheiber, R. Pippan, P. Puschnig, L. Romaner, Ab initio calculations of grain boundaries in bcc metals, 
Modelling Simul. Mater. Sci. Eng. 24 (2016) 035013.


\end{thebibliography}
\end{document}